\documentclass[11pt,twoside]{article}

\usepackage{asp2006_hotcool}
\usepackage{epsf}
\usepackage{psfig}
\usepackage{lscape}

\def\gtrsi{\mathrel{\hbox{\rlap{\hbox{\lower4pt\hbox{$\sim$}}}\hbox{$>$}}}}

\begin{document}
\title{Seeking Core-Collapse Supernova Progenitors in Pre-Explosion Images}

\author{Douglas C. Leonard}  
\affil{Department of Astronomy, San Diego State University, San Diego, CA
  92182, USA}    

\begin{abstract} 

I summarize what we have learned about the nature of stars that ultimately
explode as core-collapse supernovae from the examination of images taken prior
to the explosion.  By registering pre-supernova and post-supernova images,
usually taken at high resolution using either space-based optical detectors, or
ground-based infrared detectors equipped with laser guide star adaptive optics
systems, nearly three dozen core-collapse supernovae have now had the
properties of their progenitor stars either directly measured or (more
commonly) constrained by establishing upper limits on their luminosities.
These studies enable direct comparison with stellar evolution models that, in
turn, permit estimates of the progenitor stars' physical characteristics to be
made.  I review progenitor characteristics (or constraints) inferred from this
work for each of the major core-collapse supernova types (II-Plateau,
II-Linear, IIb, IIn, Ib/c), with a particular focus on the analytical
techniques used and the processes through which conclusions have been drawn.
Brief discussion of a few individual events is also provided, including SN
2005gl, a type IIn supernova that is shown to have had an extremely luminous --
and thus very massive -- progenitor that exploded shortly after a violent,
luminous blue variable-like eruption phase, contrary to standard theoretical
predictions.

\end{abstract}

\section{Introduction} 

This review focuses specifically on what we have learned about the progenitors
of core-collapse supernovae (CC~SNe) by examining images of the supernova (SN)
sites taken prior to the explosion.  By registering pre-SN and post-SN images,
usually taken at high resolution using either space-based optical detectors, or
ground-based infrared detectors equipped with laser guide star adaptive optics
systems (LGS-AO), about one dozen CC SN progenitors have now been directly
detected (i.e., shown to be spatially coincident with the SN) in pre-SN images,
with roughly two dozen upper limits derived from non-detections
\citep{Smartt09b}.  This field has come a long way in the last decade, and
promises to advance rapidly as more and more nearby galaxies -- hosts of future
CC SNe -- have high-resolution images added to the archive.

This review is organized as follows.  Following a brief summary of SN
classification and stellar evolution theory (\S~2), one example from each of
the following three categories of progenitor studies is provided (\S~3; ordered
from most-to-least common): (1) No progenitor star detected in pre-SN image(s);
(2) Likely progenitor star identified via spatial coincidence in pre-SN and
post-SN images; (3) Progenitor star detected in pre-SN image(s) and
subsequently confirmed by demonstrating its absence in images taken after the
SN has faded beyond detection.  A summary of overall results to date for each
SN type is then given (\S~4), followed by a brief discussion of outstanding
questions and areas in which future progress is likely (\S~5).  Note that
discussion is limited to what the examination of images of SN sites taken prior
to the explosion has taught us, and necessarily excludes (or relegates to very
brief comment) such related investigations as SN environments (e.g.,
\citealt{Vandyk99}; see also the article by Elias Rosa in this volume) and SN
progenitor ``forensics'' (e.g., \citealt{Modjaz09}; see also the article by
Modjaz in this volume).  For a comprehensive discussion of all such related
areas, see the recent review by \citet{Smartt09b}.

\section{Background:  SN Classification and Stellar Evolution}

It is typical to subdivide CC~SNe into at least five major categories (see
\citealt{Filippenko97} for a thorough review): II-Plateau (II-P; hydrogen in
spectrum and plateau in optical light curve), II-Linear (II-L; hydrogen in
spectrum, no plateau in optical light curve), IIn (hydrogen in spectrum and
spectral and photometric evidence for interaction between SN ejecta and a dense
circumstellar medium [CSM]), IIb (hydrogen in spectrum initially, but
transforms into a hydrogen-deficient spectrum at later times), and Ib/c (no
evidence for hydrogen in spectrum at any time), where the ordering is a roughly
increasing one in terms of inferred degree of envelope stripping prior to
explosion (i.e., II-P are the least stripped at the time of explosion, and Ib/c
are the most stripped).

While most of this review focuses on the observational advances that have been
made, theoretical input is critical to translate observed progenitor luminosity
(or limits) into zero-age-main-sequence masses ($M_{\rm ZAMS}$) and stellar
evolutionary states.  Among the most complete (and accessible\footnote{The
models can be downloaded from the code's Web site, at
http://www.ast.cam.ac.uk/$\sim$stars.}) stellar evolution models at present are
the metallicity-dependent models produced with the Cambridge stellar evolution
code, STARS, the descendant of the code developed originally by
\citet{Eggleton71} and updated most recently by \citeauthor{Eldridge04} (2004;
see also \citealt{Smartt09a}, and references therein), since they follow stellar
evolution up to the initiation of core neon burning, which is likely to give an
accurate indication of the pre-SN luminosity.  The Hertzsprung-Russell diagram
(HRD) of the STARS evolutionary tracks are shown in Figure~1 for stars ranging
in initial mass from $6\ M_\odot \rightarrow 100\ M_\odot$.  Comparison with
other contemporary model grids (e.g., \citealt{Heger00}; \citealt{Meynet00})
show that the endpoints for stars in the $8 \rightarrow 15 \ M_\odot$ range
differ by at most $0.2$ dex in luminosity among the codes \citep{Smartt04},
which gives some assurance that systematic uncertainties are not great, at
least at the low-mass end for red supergiant (RSG) stars.  Two areas of
uncertainty in need of better quantification (or, at least, agreement within
the community) include the effects that stellar rotation and mass-loss might
have on the observable characteristics of stars prior to core collapse.

\begin{figure}[!ht]

\plotone{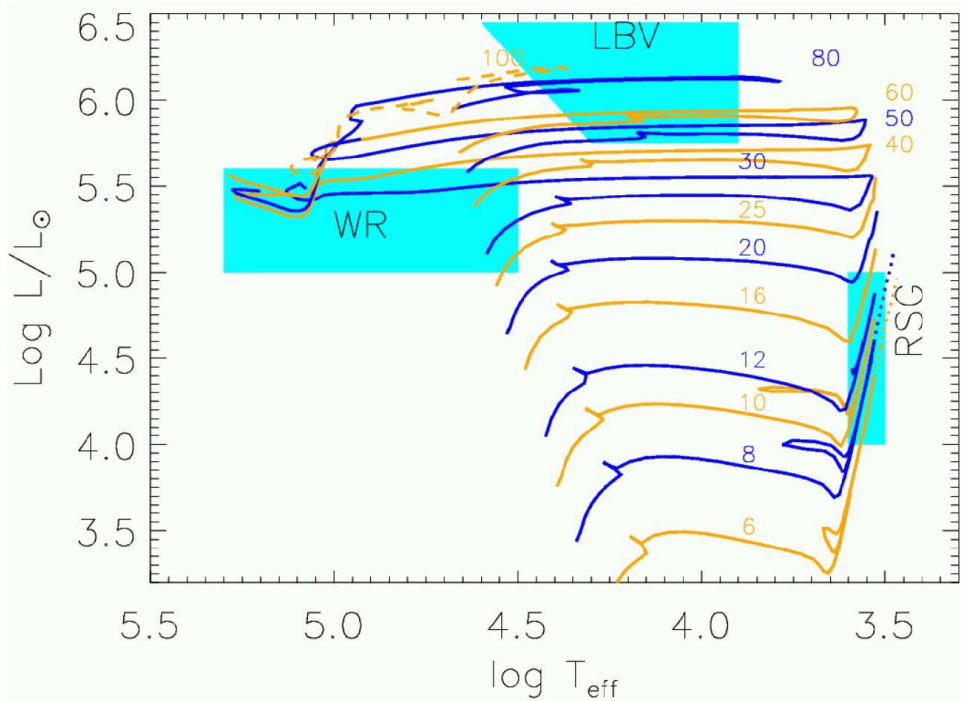}

\caption{The HRD of the STARS evolutionary tracks, from \citet{Smartt09b}.  The
  location of RSG, Wolf-Rayet (W-R), and luminous blue variable (LBV) stars are
  indicated by {\it shaded regions}. Figure reprinted, with permission, from
  the {\it Annual Review of Astronomy and Astrophysics}, Volume 47 \copyright
  2009 by Annual Reviews {\it www.annualreviews.org}. }

\end{figure}

\section{Seeking SN Progenitors in Pre-Explosion Images:  Case Studies}

\subsection{Case I: No Progenitor Detected in Pre-Explosion Images}

Not surprisingly, when no progenitor star is actually detected at the SN
location in pre-SN images, only an upper limit to the progenitor's luminosity
and, hence, mass, can be derived.  To illustrate the analysis process in such a
situation, I consider SN~2006my, an SN~II-P that exploded in a galaxy $\sim 20$
Mpc away (nearly all SNe with progenitor studies are within $\sim 30$ Mpc,
since source confusion becomes an increasing problem with distance).  Details
for this particular event are provided by \citet{Leonard18}; here I briefly
outline the steps my colleagues and I took to derive an upper mass limit on its
progenitor.

The {\it Hubble Space Telescope} ({\it HST}) imaged the site of SN~2006my using
the Wide-Field and Planetary Camera 2 (WFPC2) in 1994 (pre-SN) and again in
2007 (shortly after explosion).  We registered the two images and pinpointed
the SN location to better than 30 milli-arcsec in the pre-SN frame (Figure
2a,b).  Such fine registration allowed us to rule out a nearby point source
(source `1' in Figure 2a) as the progenitor star with greater than $96\%$
confidence.  (Note that this source had been previously identified by
\citet{Li07} as the likely progenitor based on registration with
lower-resolution ground-based optical post-SN images.)  We next set an $I$-band
detection limit in the pre-SN frame by placing artificial stars of
progressively fainter magnitude at the SN location and letting the photometry
software (in this case, HSTphot, see \citealt{Dolphin00a}) attempt to detect
them.  The point at which the software no longer detected a point source then
serves as the limiting upper magnitude for the progenitor star.

\begin{figure}[!ht]

\plotone{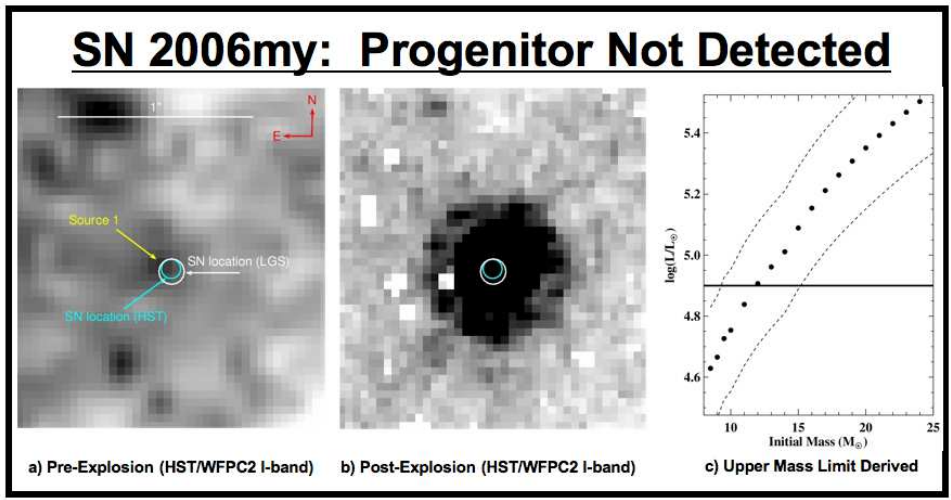}

\caption{Site of SN~2006my in pre-SN image ({\it a}) and post-SN image ({\it
  b}), both taken with {\it HST}. {\it Circles} in ({\it a}) and ({\it b})
  represent the approximate $5\sigma$ uncertainty of the position of SN~2006my
  when the pre-SN image is registered with the {\it HST} image (smaller
  circles) and a ground-based image taken with LGS-AO (slightly larger
  circles).  Source 1, labeled in pre-SN image ({\it a}), is a nearby source
  determined to {\it not} be coincident with SN~2006my at the 96\% confidence
  level.  Panel ({\it c}) presents the initial mass vs. final predicted
  luminosity prior to explosion for $Z = 0.01$ stars ({\it filled circles})
  evolved with the STARS stellar evolution code \citep{Eldridge04}.  {\it
  Dashed lines} indicate the estimated systematic uncertainty in the stars'
  final luminosities, as described in the text (see \S~2).  The {\it solid,
  horizontal line} represents the $3\sigma$ upper luminosity limit for a RSG
  that could have remained undetected by the analysis of pre-SN images of the
  site of SN~2006my.  Figure adapted from \citet{Leonard18}, and reproduced by
  kind permission of University of Chicago Press.}
\end{figure}

To translate this single-filter detection limit into a luminosity, we assumed
that the progenitor was a RSG (given other SN~II-P progenitor detections this
seems a reasonable assumption; see \S 3.2 and \S 4), and then determined the
greatest bolometric magnitude it could have had while still remaining below our
detection threshold.  This is accomplished through:

$$M_{\rm bol} = -\mu - A_V + I + (V - I)_{\rm RSG} + {\rm BC}_V,$$

\noindent where $\mu$ is the distance modulus of the host galaxy (NGC 4651),
$A_V$ the extinction to SN~2006my, $I$ the $I$-band detection threshold, $(V -
I)_{\rm RSG}$ the color range of RSG stars (i.e., spectral types ${\rm K3}
\rightarrow {\rm M4}$), and ${\rm BC}_V$ the bolometric correction
corresponding to each $(V - I)_{\rm RSG}$.  Upon adopting the most conservative
values for each of the parameters (i.e., the ones that produce the least
restrictive $M_{\rm bol}$ for the progenitor's upper luminosity limit), and
allowing for a maximum systematic uncertainty of 0.2 dex in the theoretical
stellar model endpoints (see \S~2), the limiting bolometric magnitude above
which any RSG would have been detected in our pre-SN image, $M_{\rm bol}$, is
derived.  We then compared this with the final luminosity of stars with $M_{\rm
ZAMS} > 8\ M_\odot$ predicted by the STARS stellar evolution models (Figure 2c)
to derive an upper bound on the progenitor mass of $M_{\rm ZAMS} = 15 \
M_\odot$.  From this analysis, then, we conclude that any RSG progenitor with
an initial mass greater than $15\ M_\odot$ would have been detected using our
analysis procedure.

Analyses similar to that described here for SN~2006my have been carried out on
each of 22 non-detections in pre-SN images \citep{Smartt09b}.  As we shall see
(\S~4), it is the sheer number of such progenitor non-detections that permits
rather strong conclusions to be drawn about CC SN progenitors from this
category of progenitor studies.

\subsection{Case II: Putative Progenitor Detected in Pre-Explosion Images}

Next, we consider the individually more revealing situation where an object
coincident with the transformed SN location is actually detected in the pre-SN
image(s), a situation that exists now for 11 CC SNe \citep{Smartt09b}.  As an
outstanding example of the analytic power provided by having multi-filter
pre-SN images available (especially in the near infrared for RSG progenitors),
we consider the recent work of \citet{Mattila08} on SN~2008bk, a very nearby
($\sim 4$ Mpc) SN~II-P.  In this case, pre-SN ground-based images in $BVIJHK$
were registered with post-SN LGS-AO $K$-band images to yield solid progenitor
star detections in $IJHK$, and upper luminosity limits in $B$ and $V$.  When
compared with the known spectral energy distribution (SED) of RSG, a good match
for the progenitor of SN~2008bk is found with a progenitor of spectral type M4I
(Figure~3).  From comparison with the STARS stellar evolutionary models an
initial progenitor mass of $8.5 \pm 1.0 \ M_\odot$ is derived for this SN.
Similar studies on seven other detected SN~II-P progenitors have found stars
consistent with RSG in all cases, providing nice agreement between theory and
observation.  As we shall see in \S~4, however, the {\it range} of masses
inferred for these RSG progenitors is somewhat unexpected.

\begin{figure}[!ht]

\begin{center}

\scalebox{0.7}{

\plotone{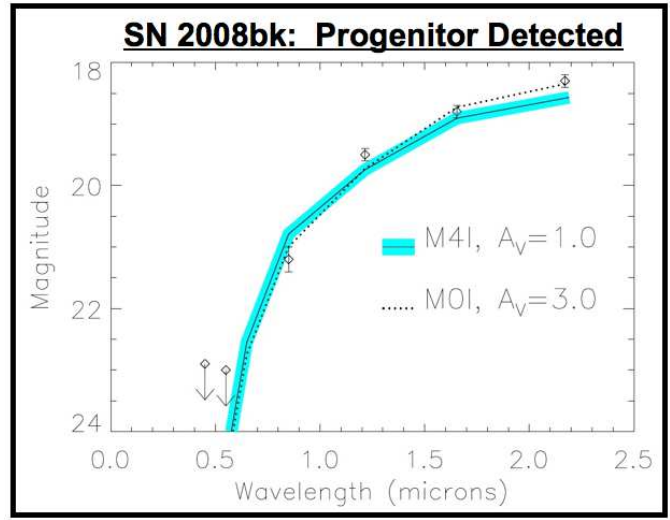}
}

\end{center}

\caption{Observed SED (open diamonds) of the detected progenitor for SN~2008bk
  compared with the reddened colors of an M4I supergiant ({\it solid, broad
  line}; $A_V = 1$ mag assumed) and an M0I supergiant ({\it dotted line}; $A_V
  = 3$ mag assumed). Since spectra of SN~2008bk reveal little evidence of
  significant reddening, the M4I supergiant progenitor is favored.  The width
  of the M4I line indicates the range in observed colors of M4I stars, from
  \citet{Elias85}.  Figure reproduced by kind permission of Blackwell
  Publishing Ltd. }

\end{figure}

\subsection{Case III: Progenitor Detected and Confirmed}

Finally, we consider the most satisfying situation where images taken before,
during, and long after the SN explosion exist that clearly show the progenitor
star, the SN, and the absence of the progenitor star, respectively.  Such a
sequence provides nearly conclusive proof of the progenitor star's
identity.\footnote{Dust obscuration remains a difficult possibility to
  definitively exclude -- e.g., if substantial dust is formed in the SN
  atmosphere {\it and} the putative progenitor star lies behind the SN along
  the l-o-s, the star could be obscured in post-SN images.}  Currently, such a
time series exists for only two objects:\footnote{``Only two objects'' as of
  February 2009, when this review was written.  SN~1993J and SN 2003gd should
  now (August 2009) properly be added to the list of proposed progenitor stars
  confirmed to have disappeared in post-SN images; see \citet{Maund09}.}
SN~1987A \citep[e.g.,][]{Graves05} and SN~2005gl \citep{Galyam07,Galyam09}.
Because the case of SN~1987A is well-known, I present SN~2005gl as the example
of this situation; it also clearly demonstrates the investigative power
provided by having a third observation, long after the SN has dropped below
detection.

As shown by \citet{Galyam07}, early spectra of SN~2005gl exhibited the classic
features of a Type IIn event, showing narrow but resolved lines of hydrogen
superposed on an intermediate-width component on an otherwise featureless
continuum.  Analysis of the spectral features indicate ejecta interacting with
a dense CSM, whose properties suggest that the progenitor star exploded shortly
after an LBV-like mass-loss episode.  Comparison of a pre-SN {\it HST} image
with a post-SN image obtained from the ground using the LGS-AO at the Keck II
telescope established a spatial coincidence between the SN and a very bright
source possessing an estimated luminosity of over $10^6\ L_\odot$
(\citealt{Galyam07}; see Figures~4a and 4b).  The only single stars known to
possess such an extraordinary luminosity are very massive
($\mathrel{\hbox{\rlap{\hbox{\lower4pt\hbox{$\sim$}}}\hbox{$>$}}} 70\ M_\odot$;
see Figure~1), which conventional theory predicts should explode only after the
LBV phase has ended \citep{Maeder94}.

Initially, strong claims for the unexpectedly luminous progenitor/SN~2005gl
association had to be tempered by consideration of the distance of SN~2005gl's
host galaxy.  At over 60 Mpc away, the $\sim 0.1\arcsec$ resolution of the
pre-SN {\it HST} image corresponds to $\sim 30$ pc, which raises suspicion that
the object could be, e.g., an unresolved stellar cluster or association of
several massive stars, with only part of the light coming from the actual
progenitor of SN~2005gl \citep{Galyam07}.  Additional observations, therefore,
were clearly needed to settle the case, and two years later, an additional {\it
HST} observation was made.  This observation demonstrates that the luminous
source in the pre-SN image has, indeed, disappeared (Figure~4c), which implies
that the progenitor of SN~2005gl was a single, extremely luminous, star that
exploded while in the LBV phase \citep{Galyam09}.  Such a luminosity is
indicative of having had an initial mass of $M_{\rm ZAMS}
\mathrel{\hbox{\rlap{\hbox{\lower4pt\hbox{$\sim$}}}\hbox{$>$}}} 70\ M_\odot$,
which likely left behind a stellar mass black hole \citep[e.g.,][]{Orosz07}.
In addition to exploding during an unexpected evolutionary phase, the very fact
{\it that} such a massive star is demonstrated to have exploded at all -- as
opposed to directly collapsing to a black hole with no SN explosion -- is
important, since the optical signature produced at the time of stellar collapse
to a black hole is, at present, virtually unconstrained by either observation
or theory \citep[see, e.g., ][and references therein]{Kochanek08}.

\begin{figure}[!ht]

\plotone{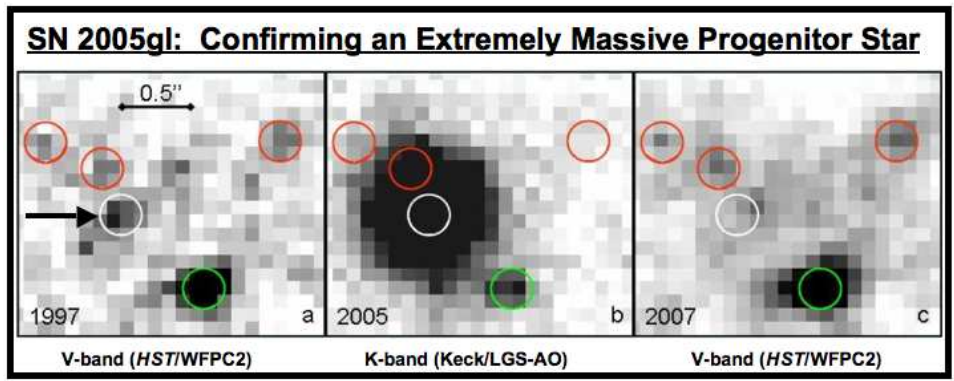}

\caption{A demonstration that the progenitor star of SN~2005gl has vanished,
  using images taken before ({\it a}, progenitor star indicated with {\it
  arrow}), shortly after ({\it b}), and long after the SN explosion ({\it c}),
  from \citet{Galyam09}.  Figure reproduced by kind permission of Nature
  Publishing Group.}

\end{figure}

\section{Summary of Results to Date}

The three examples discussed in \S~3 serve to illustrate how the science of
seeking progenitors in pre-SN images is carried out, and what conclusions can
typically be drawn.  I now briefly summarize results to date arising from
direct progenitor searches in pre-SN images; for a more comprehensive review,
see \citet{Smartt09b}.

{\bf Type II-Plateau:} SNe~II-P are by far the most well-defined category of CC
SNe in terms of direct observational progenitor constraints, having had
eight putative progenitor detections made and 12 upper luminosity limits
established.  All of the available evidence suggests that RSG are the immediate
progenitors of SNe~II-P.  By employing a uniform reduction and analysis
procedure, \citet{Smartt09b} has produced the cumulative frequency distribution
shown in Figure~5 for SNe~II-P, from which an intriguing result is immediately
evident: All but one of the SNe~II-P have initial masses constrained to be
$\mathrel{\hbox{\rlap{\hbox{\lower4pt\hbox{$\sim$}}}\hbox{$<$}}} 18\ M_\odot$,
with the most massive {\it detected} progenitor of an SN~II-P having a mass of
only $16.5\ M_\odot$.  This is surprising, since RSG up to $25\ M_\odot$ are
clearly observed in the Local Group \citep[][and references therein]{Smartt09a},
and would have easily been detected in the pre-SN images.  This lack of massive
RSG progenitors for SNe II-P lead \citet{Smartt09a} to speculate that these
massive RSG progenitors may be forming black holes heralded by faint, or
non-existent, SN explosions \citep[see also][]{Kochanek08}.

\begin{figure}[!ht]

\begin{center}

\scalebox{0.6}{

\plotone{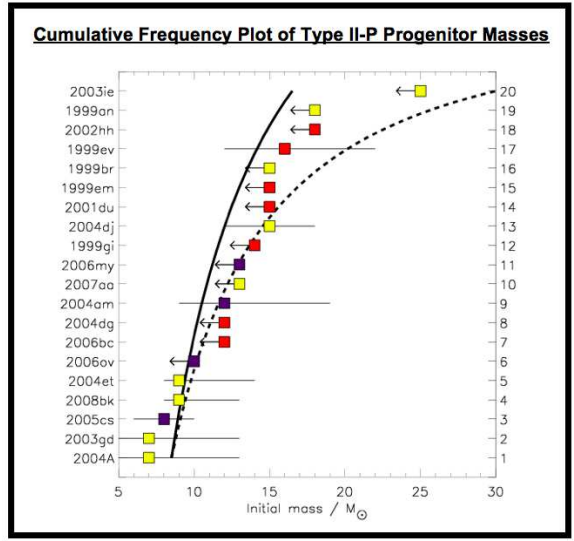}
}

\end{center}
\caption{Initial masses of SNe~II-P progenitor stars displayed as a cumulative
  frequency plot, from \citet{Smartt09b}.  Right-hand axis gives the number of
  SNe~II-P known to have progenitors less massive than the given SN.  {\it
  Solid line} is a Salpeter initial mass function (IMF; $\alpha = -2.35$) with
  a minimum mass of $8.5\ M_\odot$ and a maximum mass of $16.5\ M_\odot$, and
  represents the most likely fit to the data.  {\it Dashed line} is a Salpeter
  IMF but with a maximum mass of $30\ M_\odot$, and is a poor fit to the data.
  Figure reprinted, with permission, from the {\it Annual Review of Astronomy
  and Astrophysics}, Volume 47 \copyright 2009 by Annual Reviews {\it
  www.annualreviews.org}.}

\end{figure}

{\bf Type II-Linear:} A rare type of CC SN, it is perhaps not surprising that
only one SN~II-L (SN~1980K) has a pre-SN image, the analysis of which rules out
massive RSG greater than about $18\ M_\odot$ \citep{Thompson82}.  Analysis of
the stellar population of the Type II-L SN~1979C by \citet{Vandyk99a}
determines a mass range of $15 - 21\ M_\odot$ for its progenitor.  At
this point, firm conclusions about the progenitors of SNe~II-L can not be made,
although early indications are that at least some do not arise from extremely
massive stars.

{\bf Type IIn:} SN~2005gl, described earlier in this review (\S~3.3) as having
a very massive
($\mathrel{\hbox{\rlap{\hbox{\lower4pt\hbox{$\sim$}}}\hbox{$>$}}} 70\ M_\odot$)
progenitor that exploded while in the LBV phase, is the only example of an
SN~IIn for which a progenitor has been detected in pre-SN images.  Whether such
a massive progenitor is indicative of the class as a whole is not known.

{\bf Type IIb:} Pre-SN images exist for two events.  First, SN~1993J in M81,
where extensive analyses of pre-SN and post-SN images (and spectra) lead to the
conclusion that a $13 - 20\ M_\odot$ star exploded in a binary system, with a
slightly less massive secondary surviving the explosion \citep[][and references
therein]{Maund04}.  Very recently, the Type IIb SN~2008ax has provided a great
opportunity to further investigate this rare class of CC SNe since pre-SN {\it
HST}/WFPC2 images exist in $BVI$.  A study by \citet{Crockett08} finds a
curiously flat SED for the progenitor star, which is impossible to reconcile
with a single RSG, but may be consistent with an early-type W-R (WN class)
progenitor, suggesting a progenitor star with a large ($25 - 30\ M_\odot$)
initial mass.

{\bf Type Ib/c:} A well-studied class, with ten upper limits but no detections
from analysis of pre-SN images.  The lack of detections is surprising, since it
is commonly thought that at least some of the progenitors of SNe~Ib/c should be
luminous, single W-R stars, in addition to others perhaps arising from lower
mass stars in binary systems.  While none of the non-detections definitively
rule out a W-R progenitor, \citet{Smartt09b} demonstrates that it is quite
unlikely at this point that all SNe~Ib/c come from them.

A summary of the current state of affairs of CC SN progenitor research via
studies of pre-SN images is provided by Figure~6.

\begin{figure}[!ht]

\begin{center}
\scalebox{0.7}{
\plotone{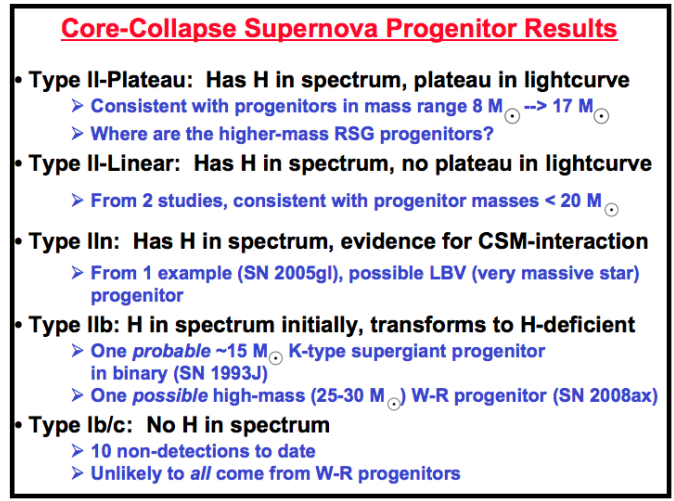}
}

\end{center}
\caption{A concise summary of CC~SN progenitor knowledge gained solely from
  investigations of pre-SN images (in the case of SNe~IIL, the post-SN image
  environment study by \citet{Vandyk99a} is also incorporated).}

\end{figure}

\section{Future Directions}

The science of seeking SN progenitors has made tremendous strides in just the
last ten years.  For the future, I look with particular interest at the
extremes as areas ripe for breakthrough discoveries, since it is there that
many of our most fundamental questions lie.  On the low-mass end, how will the
lack of massive RSG progenitors for SNe~II-P be resolved?  Are we seeing the
first glimpse of the mass cutoff for direct collapse to black holes?  If so,
then how will this be reconciled with the {\it very} massive stars (i.e., the
other mass extreme) that apparently do explode as SNe~IIn or possibly IIb?  And
finally, how does binarity influence all of these conclusions?  Clearly, we are
just at the beginning stages of this exciting field of research, and great
advances will no doubt be made in the coming decade.

\acknowledgements 

I thank the scientific organizing committee of the ``Hot And Cool: Bridging
Gaps in Massive Star Evolution'' conference for inviting me to provide this
review. I thank Seppo Mattila for permitting reproduction of a figure from a
recent paper, and Stephen Smartt for allowing the use of figures in advance of
publication of his Annual Reviews article on the topic.

\end{document}